\RequirePackage{fix-cm}
\documentclass[smallextended]{svjour3}       
\smartqed  
\usepackage{graphicx}
\begin{document}

\title{Charged Polytropic Models on Finch-Skea Spacetime
}


\author{B. S. Ratanpal}


\institute{B. S. Ratanpal \at
              Department of Applied Mathematics, Faculty of Technology \& Engineering, The Maharaja Sayajirao University of Baroda, Vadodara - 390001, India \\
              \email{bharatratanpal@gmail.com}           
}

\date{Received: date / Accepted: date}

\maketitle

\begin{abstract}
The static, charged, spherically symmetric matter distribution have been studied by considering polytropic equation of state. Two polytropic indices have been considered for study. The plots of density, radial pressure, tangential pressure, anisotropy and causality condition indicates the models are suitable to describe the relativistic polytropes.
\end{abstract}

\section{Introduction}
\label{sec:1}
The Lane-Emden equation plays the important role to model the compact stellar structures. The solution of Lane-Emden equation describes polytropes. The polytropic equation of state is useful to study the fundamental problems in astrophysics(\cite{GW},\cite{A1}). Newtonian polytropes were studied by Chandrasekhar\cite{C1}. Kovetz\cite{K1} studied slowly rotating polytropes using Chandrasekhar's work. The models decribing polytropes in general relativity were developed by Azar \textit{et. al.}\cite{AMNR}, Herrera \& Berreto\cite{HB} and Topper(\cite{T1}\cite{T2}\cite{T3}). Theoretical investigation of Ruderman\cite{R1} concludes that when the matter density is much higher than nuclear density, pressure may not be isotropic in nature. The effect of anisotropy on surface redshift can be found in the work of Bowers \& Liang\cite{BL} and Herrera \& Santos\cite{HS}. Anisotropy can occur due to type-3A super fluid\cite{KW} or phase transition\cite{S1}. The algorithm to find anisotropic solutions from known isotropic solutions have been developed by Maharaj and Chaisi\cite{MC}. The anisotropic stars with linear equation of state have been studied by Sharma and Maharaj\cite{SM}. Sharma and Ratanpal\cite{SR} studied anisotropic models of superdens stars admitting quadratic equation of state. The charged dust models in equilibrium with uniform density have been studied by Cooperstock and de la Cruz\cite{CC}. Stellar models with radially increasing matter density have been studied by Bonner and Wickramasuriya\cite{BW}. Pant and Sah\cite{PS} derived exact solution for charged distribution. The necessary conditions for solution of Einstein-Maxwel system of equations to be regular inside the stellar object have been obtained by Maartens and Maharaj\cite{MM}. Several authors have obtained analytic solution of Einstein-Maxwel system of equation including Ratanpal \textit{et. al.}\cite{RPSD}, Ratanpal \& Sharma\cite{RS}, Thomas \& Pandya\cite{TP}, Ratanpal \textit{et. al.}\cite{RTP}, Patel \& Kopper\cite{PK}, Komathiraj \& Maharaj\cite{KM} and Thirukkanesh and Maharaj\cite{TM1}.\\\\
In this paper charged anisotropic model of stellar configuration satisfying polytropic equation of state have been studied on Finch-Skea\cite{FS} spacetime. The charged anisotropic model satisfying polytropic equation of state is also studied by Takisa and Maharaj\cite{TM2}, by choosing metric potential $g_{rr}=\frac{1+ar^2}{1+br^2}$, if $b=0$ then this metric potential reduces to $g_{rr}=1+ar^2$ which appears in Finch-Skea\cite{FS} spacetime. However, the equations governing density, pressure etc. in Takisa and Maharaj\cite{TM2} model contains the parameter $b$ in the denominator. Hence our model is not a particular case of Takisa and Maharaj\cite{TM2} model. In sect. 2, Einstein-Maxwell system is described for particular form of electric field intensity. The Einstein-Maxwell system is then integrated in Sect. 3 by considering two particular values of polytropic index $\Gamma$. The physical features of the model have been discussed in sect. 4. Sect. 5 contains the discussion.
\section{Einstein-Maxwell System of Equations}
\label{sec:2}
We consider static spherically symmetric spacetime metric
\begin{equation}\label{M1}
	ds^2=e^{\nu(r)}dt^2-e^{\lambda(r)}dr^2-r^2\left(d\theta^2+\sin^2\theta d\phi^2\right),
\end{equation}
with energy-mommentum tensor for anisotropic charged distribution as 
\begin{equation}\label{EMTensor}
	T_{ij}=diag\left(\rho+E^2,\;p_{r}-E^2,\;p_{\perp}+E^2,\;p_{\perp}+E^2\right),
\end{equation}
where $\rho$ represents energy density, $p_{r}$ represents radial pressure, $p_{\perp}$ represents tangential presure and $E$ represents energy density. These quantities are measured relative to unit four velocity $u^{i}=e^{\nu}\delta_{0}^{i}$. The Einstein-Maxwell system is then written as 
\begin{equation}\label{FE1}
	8\pi\rho+E^2=\frac{1-e^{-\lambda}}{r^2}+\frac{\lambda' e^{-\lambda}}{r},
\end{equation}
\begin{equation}\label{FE2}
	8\pi p_{r}-E^2=\frac{e^{-\lambda}-1}{r^2}+\frac{e^{-\lambda}\nu'}{r},
\end{equation}
\begin{equation}\label{FE3}
	8\pi p_{\perp}+E^2=e^{-\lambda}\left(\frac{\nu''}{2}+\frac{\nu'^2}{4}-\frac{\nu'\lambda'}{4}+\frac{\nu'-\lambda'}{2r} \right),
\end{equation}
\begin{equation}\label{FE4}
	q(r)=4\pi\int_{0}^{r} \sigma r^2 e^{\lambda/2}dr,
\end{equation}
and electric field intensity is given by
\begin{equation}\label{EFI}
	E=\frac{q(r)}{r^2}.
\end{equation}
We make the choice
\begin{equation}\label{eLam}
	e^{\lambda}=1+ar^2,
\end{equation}
with this ansatz spacetime metric (\ref{M1}) reduces to Finch-Skea\cite{FS} spacetime metric and we define electric field intensity as 
\begin{equation}\label{E}
	E^2=\frac{\alpha r^2}{\left(1+ar^2 \right)^2}.
\end{equation}
This choise of electric field intensity is physically acceptable as it is zero at the centre, finite throughout the distribution. We define anisotropy as 
\begin{equation}\label{S}
	\Delta=p_{\perp}-p_{r},
\end{equation}
using (\ref{eLam}) and (\ref{E}) in (\ref{FE1}) density takes the form
\begin{equation}\label{rho1}
	8\pi\rho=\frac{3a+\left(a^2-\alpha \right)r^2}{\left(1+ar^2 \right)^2},
\end{equation}
which is finite at centre of the stellar configuration. We consider the polytropic equation of state to model the stellar configuration i.e.
\begin{equation}\label{pr1}
	p_{r}=K\rho^{\Gamma},
\end{equation}
where $\Gamma=1+\frac{1}{\eta}$. From (\ref{rho1}) and (\ref{pr1}) the equation of radial pressure takes the form
\begin{equation}\label{pr2}
	8\pi p_{r}=K\left[\frac{3a+\left(a^2-\alpha \right)}{\left(1+ar^2 \right)^2} \right]^\Gamma,
\end{equation}
using (\ref{eLam}), (\ref{E}) and (\ref{pr2}) in (\ref{FE2}) we get
\begin{equation}\label{nudash}
	\nu'=\frac{rK\left[3a+\left(a^2-\alpha \right)r^2 \right]^\Gamma}{\left(1+ar^2 \right)^{2\Gamma -1}}+\frac{r\left[a+\left(a^2-\alpha \right)r^2 \right]}{\left(1+ar^2\right)},
\end{equation}
equation (\ref{nudash}) have been integrated in next section for two particular choices of $\eta$ and hence $\Gamma$.

\section{Solution of Einstein-Maxwell System of equations}
\label{sec:3}
Takisa and Maharaj\cite{TM2} solved the Einstein-Maxwell system of equations for polytropic star by considering $\eta=1,\;2,\;\frac{2}{3},\;\frac{1}{2}$. In this work Einstein-Maxwell system of equations (\ref{FE1}) - (\ref{FE3}) for stellar configuration following polytropic equation of state have been integrated for $\eta=1$ and $\eta=2$.
\subsection{Case-I, $\eta=1$:}
\label{sec:3.1}
With $\eta=1$, the equation of state (\ref{pr1}) takes the form
\begin{equation}\label{pr2}
	p_{r}=K\rho^2,
\end{equation}
and integrating (\ref{nudash}) we get
\begin{equation}\label{nu}
	\nu=logA\left(1+ar^2\right)^B+C(r),
\end{equation}
where $A$ is constant of integration,
\begin{equation}\label{B}
	B=\frac{a^{4}K+a\alpha-2a^{2}K\alpha+K\alpha^{2}}{2a^{3}},
\end{equation}
\begin{equation}\label{Cr}
	C(r)=\frac{2a^{2}r^{2}\left(a^{2}-\alpha \right)\left(1+ar^2 \right)^2-K\left(2a^{2}+\alpha \right)^2+4K\left(-2a^{4}+a^{2}\alpha+\alpha^{2} \right)\left(1+ar^2\right)}{4a^{3}\left(1+ar^2\right)^2},
\end{equation}
hence spacetime metric (\ref{M1}) takes the form
\begin{equation}\label{M2}
	ds^2=A\left(1+ar^2 \right)^B e^{C(r)}dt^2-\left(1+ar^2 \right)dr^2-r^2\left(d\theta^2+\sin^2\theta d\phi^2 \right).
\end{equation}
The expressions of density, radial pressure, tangential pressure and anisotropy respectively takes the form
\begin{equation}\label{rho3}
	8\pi\rho=\frac{3a+\left(a^2-\alpha\right)r^2}{\left(1+ar^2\right)^2},
\end{equation}
\begin{equation}\label{pr3}
	8\pi p_{r}=\frac{K\left[3a+\left(a^2 -\alpha \right)r^2 \right]^2}{\left(1+ar^2\right)^4},
\end{equation}
\begin{equation}\label{pp3}
	8\pi p_{\perp}=\frac{X_{1}(r)-X_{2}(r)+X_{3}(r)-X_{4}(r)+X_{5}(r)+X_{6}(r)-X_{7}(r)}{4\left(1+ar^2 \right)^7},
\end{equation}
where,\\
$X_{1}(r)=r^2\left\{\left(1+ar^2 \right)^2 \left[a+r^2\left(a^2-\alpha \right) \right]+K\left[3a+r^2\left(a^2-\alpha\right) \right]^2 \right\}^2$,\\
$X_{2}(r)=6ar^2\left[a+r^2\left(a^2-\alpha \right) \right]\left(1+ar^2 \right)^4$,\\
$X_{3}(r)=4K\left[3a+r^2\left(a^2 -\alpha\right) \right]^{2}\left(1+ar^2 \right)^{3}$,\\
$X_{4}(r)=14aK\left[3ar+r^3\left(a^2 -\alpha \right) \right]^{2}\left(1+ar^2 \right)^{2}$,\\
$X_{5}(r)=8r^2 \left(a^2 -\alpha\right) \left(1+ar^2 \right)^5$,\\
$X_{6}(r)=8Kr^2 \left[3a+r^2 \left(a^2-\alpha \right) \right]\left(a^2-\alpha\right)\left(1+ar^2 \right)^3$,\\
$X_{7}(r)=4r^2 \alpha \left(1+ar^2 \right)^5$,\\
and 
\begin{equation}\label{S3}
	\Delta=\frac{-X_{1}(r)+X_{2}(r)+X_{4}(r)-X_{5}(r)-X_{6}(r)+X_{7}(r)}{4\left(1+ar^2 \right)^7.}
\end{equation}
Anisotropy $\Delta=0$ at $r=0$.
\subsection{Case-II, $\eta=2$:}
\label{sec:3.2}
With $\eta=2$, the equation of state (\ref{pr1}) takes the form
\begin{equation}\label{pr3}
	pr=K\rho^{3/2},
\end{equation}
and integrating (\ref{nudash}) we get
\begin{equation}\label{nu1}
	\nu=logD\left(1+ar^2 \right)^{\left(\alpha/2a^2 \right)}+E(r),
\end{equation}
where $D$ is constant of integration and 
\begin{equation}\label{Er}
	E(r)=\frac{F_{1}(r)-F_{2}(r)}{2a^{5/2}\left(1+ar^2 \right)},
\end{equation}
where,\\
$F_{1}(r)=a^{3/2}r^2 \left(a^2-\alpha \right)\left(1+ar^2 \right)+K\sqrt{a}\sqrt{3a+a^2 r^2-r^2 \alpha}\left(2a^3 r^2 -3\alpha-2ar^2 \alpha \right)$,\\
$F_{2}(r)=3K\left(a^2-\alpha \right)\sqrt{2a^2+\alpha}\left(1+ar^2 \right)\tanh^{-1}\left[\frac{\sqrt{a}\sqrt{3a+a^2 r^2-r^2 \alpha}}{\sqrt{2a^2+\alpha}} \right]$.\\
Hence spacetime metric (\ref{M1}) takes the form
\begin{equation}\label{M3}
	ds^2=D\left(1+ar^2 \right)^{\alpha/2a^2}e^{E(r)}dt^2-\left(1+ar^2\right)dr^2-r^2\left(d\theta^2+\sin^2 \theta d\phi^2 \right).
\end{equation}
The expression of density, radial pressure, tangential pressure and anisotropy respectively takes the form
\begin{equation}\label{rho4}
	8\pi\rho=\frac{3a+\left(a^2-\alpha \right)r^2}{\left(1+ar^2 \right)^2},
\end{equation}
\begin{equation}\label{pr4}
	8\pi p_{r}=\frac{K\left[3a+r^2\left(a^2-\alpha \right) \right]^{3/2}}{\left(1+ar^2 \right)^3},
\end{equation}
\begin{equation}\label{pp4}
	8\pi p_{\perp}=\frac{-2ar^2\left(1+ar^2 \right)Y_{1}(r)+r^2 Y_{1}(r)^2-Y_{2}(r)+Y_{3}(r)+Y_{4}(r)}{4\left(1+ar^2 \right)^5},
\end{equation}
where,\\
$Y_{1}(r)=\left(1+ar^2 \right)\left\{a+r^2\left(a^2-\alpha \right)+K\left[3a+r^2\left(a^2-\alpha \right) \right]^{3/2} \right\}$,\\
$Y_{2}(r)=4\alpha r^2 \left(1+ar^2 \right)^3$,\\
$Y_{3}(r)=2\left(1+ar^2 \right)^2 \left[-a+a^3 r^4 -\alpha r^2 -a\alpha r^4 +K\left(3a+a^2r^2-r^2 \alpha \right)^{3/2} \right]$,\\
\tiny
$Y_{4}(r)=2\left(1+ar^2 \right)\left[3a^3r^4+a^4 r^6+a\left(1-4\alpha r^4 + 3K\sqrt{3a+a^2r^2-r^2\alpha} \right)-r^2 \alpha \left(3+4K\sqrt{3a+a^2r^2-r^2\alpha} \right)-a^2 r^2 \left(-3 +\alpha r^4 +5K \sqrt{3a+a^2 r^2-r^2 \alpha} \right) \right]$,\\
\normalsize 
and 
\begin{equation}\label{S4}
	\Delta=\frac{4K\left[3a+r^2\left(a^2-\alpha \right) \right]^{3/2}\left(1+ar^2 \right)^2+2ar^2\left(1+ar^2 \right)Y_{1}(r)-r^2Y_{1}(r)^2+Y_{2}(r)-Y_{3}(r)-Y_{4}(r)}{4\left(1+ar^2 \right)^5},
\end{equation}
Anisotropy $\Delta=0$ at $r=0$.
\section{Physical Plausibility Conditions}
\label{sec:4}
The two new solutions of Einstein-Maxwell system discussed in this work satisfies the physical plausibility conditions. Following Takisa and Maharaj\cite{TM2}, the values of $a,\; \alpha$ are taken as $a=5.5$ and $\alpha=1$. The value of density at $r=4$ is almost zero hence $r=4$ is considered as the radius of stellar configuration. 
\begin{figure}[h]
  \includegraphics{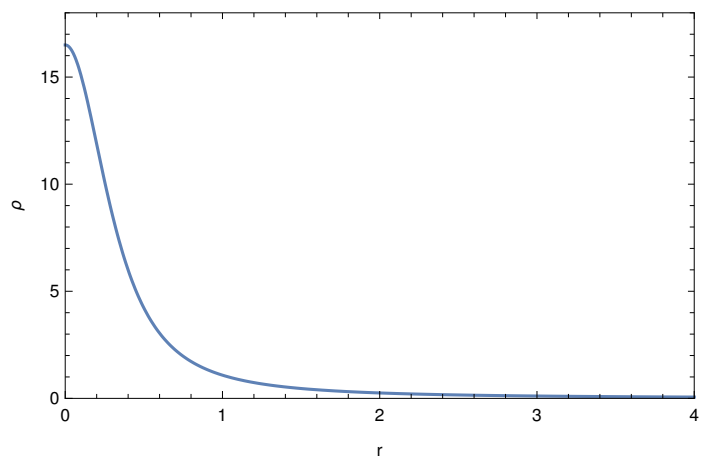}
\caption{Variation of density against r}
\label{fig1}       
\end{figure}
Fig.1 shows that $\rho$ is decreasing in radially outward direction. The condition $\frac{dp_{r}}{d\rho}<1$ gives the bound on $K$. For $\eta=1$ and $\eta=2$ the common bound found to be $K<0.0303$. To plot the graphs of physical quantities for $\eta=1$ and $\eta=2$ the value of $K$ is taken s $K=0.025$.
\begin{figure}[h]
  \includegraphics{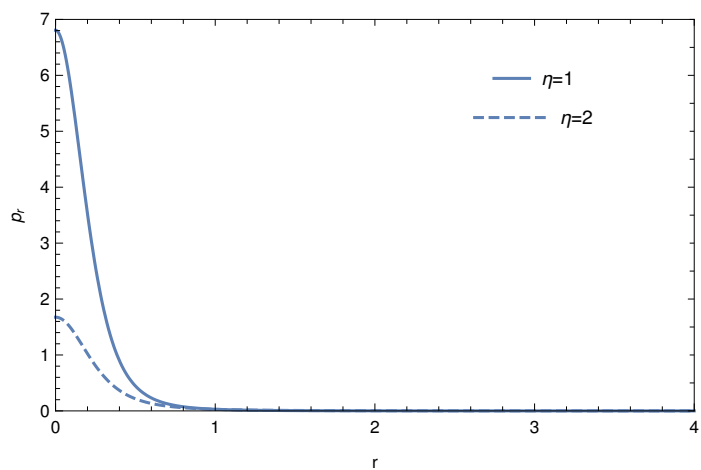}
\caption{Variation of radial pressure against r}
\label{fig2}       
\end{figure}
Fig.2 shows the graph of $p_{r}$ for $\eta=1$ and $\eta=2$. It can be observed that value of radial pressure is larger at centre of stellar configuration for $\eta=1$ compare to $\eta=2$. The radial pressure has fast decline for $\eta=1$, between $r=1$ and $r=4$ the radial pressure for $\eta=1$ and $\eta=2$ is near to 0, at $r=4$ the value of radial pressure is $p_{r}=0.0000935336$. The detailed explaination of such phenomenon can be found in Lattimer and Prakash\cite{LP}. 
\begin{figure}[h]
  \includegraphics{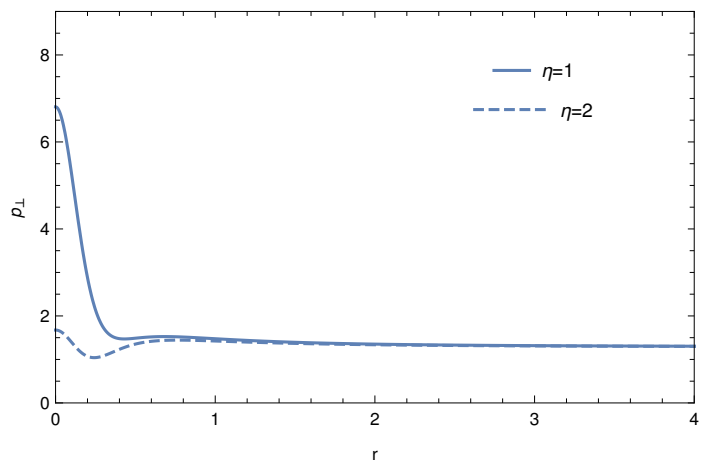}
\caption{Variation of tangential pressure against r}
\label{fig3}       
\end{figure}
The graphs of tangential pressure for $\eta=1$ and $\eta=2$ are shown in fig.3, the tangential pressure decreases than increases and take constant value, such increase in tangential pressure are available in literature (Takisa and Maharaj\cite{TM2} and Karmakar \textit{et. al.}\cite{KMSM}).
\begin{figure}[h]
  \includegraphics{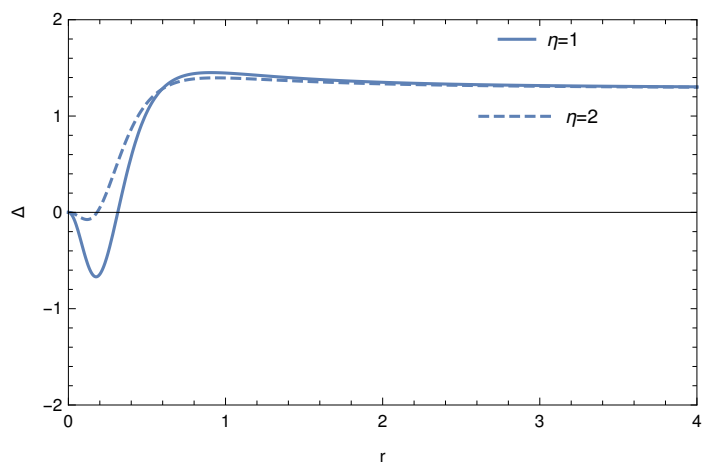}
\caption{Variation of anisotropy against r}
\label{fig4}       
\end{figure}
The variation of anisotropy $\Delta$ against radius is shown in fig.4, the anisotropy decreases than increases and constant between $r=3$ and $r=4$.
\begin{figure}[h]
  \includegraphics{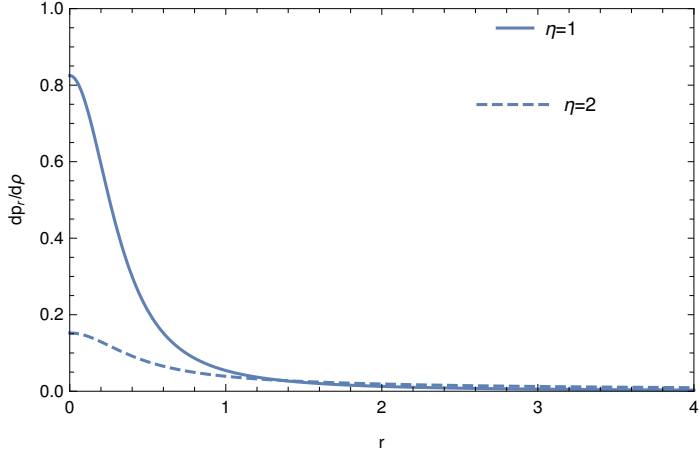}
\caption{Variation of $\frac{dp_{r}}{d\rho}$ against r}
\label{fig5}       
\end{figure}
From fig.5 it can be observed that $\frac{dp_{r}}{d\rho}<1$ for $\eta=1$ and $\eta=2$,hence causality condition is satisfied. 
\section{Discussion}
\label{sec:5}
In this work two new solution of Einstein-Maxwell system have been studied by considering polytropic equation of state. The electric field intensity is regular at the centre and well behaved throughout the stellar configuration. The obtained metric potentials $e^{\nu(r)}$ for the polytropic indices are regular. It can be observed that if we set $b=0$ in Takisa and Maharaj model\cite{TM2}, the model presented here is not a particular case of their work. the interesting feature of model is very low radial pressure between $r=1$ amd $r=4$ while density and tangential pressure are not negligible. For some regions $\frac{dp_{\perp}}{d\rho}<1$ is not satisfied. The quantity $\frac{dp}{d\rho}$ does not represent speed of sound, this was claimed by Caporaso and Brecher\cite{CB}, however this argument is controversial and not accpeted by many authors (Glass\cite{G1}). Hence model presented here satisfies the physical plausibility condition except $\frac{dp_{\perp}}{d\rho}<1$  and represents reasonably good polytropic model of relativistic star.

\begin{acknowledgements}
BSR would like to thank IUCAA, Pune for the facilities and hospitality provided to him where the part of work was carried out.
\end{acknowledgements}

%
%


\begin{thebibliography}{}
\bibitem{GW} Goldreich P. and Weber S., Astrophys. J., \textbf{238}, 991, (1980).
\bibitem{A1} Abramowicz M. A., Acta Astronaut., \textbf{33}, 313, (1983).
\bibitem{C1} Chandrasekhar S., An Introduction to the study of stellar structure (University of Chicago, Chicago, 1939).
\bibitem{K1} Kovetz A., Astrophys. J., \textbf{154}, 999, (1968).
\bibitem{AMNR} Azam M., Mardan S. A., Noureen I., and Rehman M. A., Eur. Phys. J. C., \textbf{76}, 315, (2016).
\bibitem{HB} Herrera L. and Barreto W., Gen. Rel. Grav., \textbf{36}, 127, (2004).
\bibitem{T1} Tooper R., Astrophys. J., \textbf{140}, 434, (1964).
\bibitem{T2} Tooper R., Astrophys. J., \textbf{142}, 1541, (1965).
\bibitem{T3} Tooper R., Astrophys. J., \textbf{143}, 465, (1966).
\bibitem{R1} Ruderman R., Annu. Rev. Astron. Astrophys., \textbf{10}, 427, (1972).
\bibitem{BL} Bowers R. L. and Liang E. P. T., jAstrophys. J., \textbf{188}, 53, (1997).
\bibitem{HS} Herrera L. and Santos N. O., Phys. Rep., \textbf{286}, 53, (1997).
\bibitem{KW} Kippenhahn R. and Weigert A, Stellar structure and evolution (Springer Verlang, Berlin, Heidelberg, New York, 1990)
\bibitem{S1} Sokolov A. I., J. Exp. Theoret. Phys., \textbf{52}, 575, (1980).
\bibitem{MC} Maharaj S. D. and Chaisi M., Math. Methods. Appl. Sci., \textbf{29}, 67, (2006).
\bibitem{SM} Sharma R. and Maharaj S. D., Mon. Not. R. Astron. Soc., \textbf{375}, 1265, (2007).
\bibitem{SR} Sharma R. and Ratanpal B. S., Int. J. Mod. Phys. D., \textbf{22}, 1350074-1, (2013).
\bibitem{CC} Cooperstock F. I. and de La Cruz V., Gen. Rel., Grav., \textbf{9}, 835, (1978).
\bibitem{BW} Bonner W. B. and Wickramasuriya S. B. P., Mon. Not. R. Astron. Soc., \textbf{170}, 643, (1975).
\bibitem{PS} Pant D. N. and Sah A. J. Math. Phys., \textbf{20}, 2537 (1971).
\bibitem{MM} Maartens R. and Maharaj S. D., J. Math. Phys., \textbf{31}, 151, (1990).
\bibitem{RPSD} Ratanpal B. S., Pandya D. M., Sharma R. and Das S., Astrophy. Space Sci., \textbf{362}, 82, (2017).
\bibitem{RS} Ratanpal B. S. and Sharma J., Pramana J. Phys., \textbf{86}, 527, (2016).
\bibitem{TP} Thomas V. O. and Pandya D. M., Astrophys. Space Sci., \textbf{360}, 39, (2015).
\bibitem{RTP} Ratanpal B. S., Thomas V. O. and Pandya D. M., Astrophys. Space Sci., \textbf{360}, 53, (2015).
\bibitem{PK} Patel L. K. and Koppar S. S., Ast. J. Phys., \textbf{40}, 441, (1987).
\bibitem{KM} Komathiraj K. and Maharaj S. D., Int. J. Mod. Phys. D., \textbf{16}, 1803, (2007).
\bibitem{TM1} Thirukkanesh S. and Maharaj S. D., Class. Quant. Grav., \textbf{25}, 235001-1, (2008).
\bibitem{FS} Finch M. R. and Skea J. E. F., Class. Quant. Grav., \textbf{6}, 467, (1989).
\bibitem{TM2} Takisa P. M. and Maharaj S. D., Gen. Rel. Grav., \textbf{45}, 1951, (2013).
\bibitem{LP} Lattimer J. M. and Prakash M., The Astrophy. J., \textbf{550}, 426, (2001).
\bibitem{KMSM} Karmakar S., Mukherjee S., Sharma R. and Maharaj S. D., Pramana J. Phys., \textbf{68}, 881, (2007).
\bibitem{CB} Caporaso G. and Brecher K., Phys. Rev. D., \textbf{20}, 1823, (1979).
\bibitem{G1} Glass E. N., Phys. Rev. D., \textbf{28}, 2693, (1983). 
\end{thebibliography}


\end{document}